\documentclass[%
 reprint,
superscriptaddress,
 amsmath,amssymb,
 aps,twocolumn,
pra,reprint,floatfix
]{revtex4-2}

\usepackage{graphicx}% Include figure files
\usepackage{dcolumn}% Align table columns on decimal point
\usepackage{bm}% bold math
\usepackage{dsfont}
\usepackage{blindtext}
\usepackage{mathtools}
\usepackage{upgreek}
\newcommand\varpm{\mathbin{\vcenter{\hbox{%
  \oalign{\hfil$\scriptstyle+$\hfil\cr
          \noalign{\kern-.3ex}
          $\scriptscriptstyle({-})$\cr}%
}}}}

\begin{document}

%\preprint{APS/123-QED}

\title{Non-Markovianity-based ultrasensitive parameter estimation}

\author{Olli Siltanen}
\email{olmisi@utu.fi}
\affiliation{Department of Mechanical and Materials Engineering, University of Turku, FI-20014 Turun yliopisto, Finland}
\affiliation{Department of Physics and Astronomy, University of Turku, FI-20014 Turun yliopisto, Finland}

\date{\today}

\begin{abstract}

Accurate parameter estimation is a central task in quantum metrology and sensing, where quantum resources can provide precision beyond classical limits. In realistic settings, however, system-environment interactions lead to decoherence, reducing these strategies to their classical counterparts. Noise is typically classified as Markovian or non-Markovian, with the latter often preserving quantum coherence longer and thus supporting better metrological performance. Still, the absence of noise is generally considered ideal. In this work, we uncover a striking reversal: certain non-Markovian environments not only outperform Markovian ones---including their quantum Cram\'er-Rao bounds---but can also surpass the entirely noiseless case. We demonstrate these findings numerically for an all-optical setup, which is experimentally feasible and can be extended to other physical platforms. In general, our results open new avenues for noise-assisted quantum metrology beyond conventional limits.

\end{abstract}

\maketitle

%\tableofcontents

%%%%%%%%%%
%%%%%%%%%%

\section{\label{intro}Introduction}

Sensitive measurements lie at the heart of scientific progress, shaping our understanding of nature~\cite{MM1887,Nagel2015,Parker2018} and driving technological advancements~\cite{Zabow2008,Shin2016,Grotti2018,Na2020,Lenk2023,Roslund2024}. Quantum sensing allows for the detection of extremely subtle signals and phenomena that are beyond the reach of classical measurement techniques. For example, quantum squeezing and Bose–Einstein condensates enable ultrasensitive gravitational and inertial sensing~\cite{McCuller2020,Szigeti2020}, while superconducting quantum interference devices (SQUIDs) and nitrogen-vacancy centers in diamonds allow for extraordinary precision in magnetic-field measurements~\cite{Greenberg1998,Rovny2022}. These methods, however, are sensitive not only to the quantity of interest but also to environmental interactions, i.e., noise.

Realistic quantum systems are open, i.e., they continuously interact with their environment~\cite{Zurek1991,Breuer2007,Rotter2015}. These interactions cause decoherence, leading to loss of quantum properties and deterioration of applications that rely on them, e.g., quantum sensing. Dephasing, in particular, plays a central role in determining the performance of quantum sensors, as it directly affects the phase information in which most parameters of interest are encoded~\cite{Huelga1997,Escher2011,Dorner2009}; in dephasing, the coherences of an open quantum system leak into the environment while its populations remain unchanged, marking pure quantum-to-classical transition~\cite{Breuer2007}. Some of the coherences can, however, occasionally flow back into the system. Such information backflow is often associated with structured environments and non-Markovian memory effects~\cite{Huelga2013,Breuer2016,Hsieh2019,Lu2020,Siltanen2021}.

%Noisy, non-Markovian parameter estimation schemes have recently attracted a lot of attention. Theoretically, non-Markovian dynamics have been shown to recover quantum Fisher information lost to decoherence~\cite{Chin2012,Smirne2016,Haase2018}, allowing estimation precisions to surpass their Markovian counterparts. Experimentally, such effects have been demonstrated in photonic, trapped-ion, and solid-state platforms, including nitrogen-vacancy centers and superconducting qubits, where reservoir engineering or feedback control produces controllable memory effects~\cite{Liu2011,Bernardes2016,Duan2018,Li2020,Guo2021}. While non-Markovian environments can indeed outperform Markovian ones, the prevailing view remains that it would be best not to have any noise at all---an assumption that we challenge in this work.

Noisy, non-Markovian parameter estimation schemes have recently attracted a lot of attention. Theoretically, non-Markovian dynamics have been shown to recover quantum Fisher information lost to decoherence, allowing estimation precisions beyond the Markovian limit~\cite{Chin2012,Berrada2013,Wang2017,Bai2019}. More general frameworks for non-Markovian metrology have also been developed~\cite{Altherr2021,Abiuso2023}. Experimentally, the non-Markovian advantage has been demonstrated, e.g., in nuclear magnetic resonance (NMR) metrology under engineered $1/f$-type noise~\cite{Yang2024} and in optical phase estimation revived by correlated Gaussian noise~\cite{Tang2025}. Non-Markovian thermometry studies have likewise reported enhanced precision compared to the Markovian case~\cite{Zhang2021,Aiache2024}. Yet, the prevailing view remains that it would be best not to have any noise at all---an assumption that we challenge in this work.

In this article, we propose an ultrasensitive two-qubit parameter estimation protocol \textit{based on} (non-Markovian) dephasing noise. We encode the parameter of interest in an ancillary qubit with Markovian dephasing. Then, we enhance the estimate's precision by post-selecting a system qubit undergoing non-Markovian dephasing. The protocol can be implemented in a linear optical setup. Here, the parameter of interest is the path difference of an unbalanced Mach-Zehnder interferometer, the open quantum system is the polarization degree of freedom of a single photon, the environment is its frequency, and the two interact in birefringent crystals, leading to polarization dephasing. With the example parameters used, the non-Markovian protocol outperforms its Markovian counterpart by several orders of magnitude, surpassing even the latter’s quantum Cram\'er-Rao bound (QCRB)~\cite{Helstrom1967} and the entirely dephasing-free case by one order of magnitude. The protocol is readily implementable with existing optical setups routinely used for open-system simulations~\cite{lin_opt_1,lin_opt_2,lin_opt_3,lin_opt_4,lin_opt_5,lin_opt_6,lin_opt_7,lin_opt_8,lin_opt_9}. Furthermore, we expect the protocol to be adaptable to a wide range of other physical platforms.

This paper is organized as follows. In Section~\ref{background}, we go through the necessary background of parameter estimation, open quantum systems, and non-Markovianity. In Section~\ref{protocol}, we describe our protocol and its linear-optical implementation. Simulation results with fixed example parameters are presented in Section~\ref{results}. Section~\ref{discussion} concludes the paper.

%\clearpage

\section{\label{background}Theoretical background}

\subsection{\label{estimation}Parameter estimation and sensitivity}

In sensitivity analysis, one is often interested in how some $\lambda$-dependent observable $O(\lambda)$ changes with small changes of $\lambda$. Here, $\lambda$ is the parameter being estimated. For small enough changes $|\delta\lambda|\ll1$, we can write
\begin{equation}
\langle O(\lambda+\delta\lambda)\rangle\approx\langle O(\lambda)\rangle+\frac{\partial\langle O(\lambda)\rangle}{\partial\lambda}\delta\lambda,
\label{sens1}
\end{equation}
where $\langle O\rangle=\text{tr}[O\rho]$ is the observable's expectation value when the system is in the state $\rho$. The difference $|\langle O(\lambda+\delta\lambda)\rangle-\langle O(\lambda)\rangle|$ can be experimentally detected if it is greater than the standard deviation $\Delta O:=\sqrt{\langle O^2\rangle-\langle O\rangle^2}$; any smaller difference could be interpreted as a result of the observable's intrinsic spread. Thus, $\delta\lambda$ needs to satisfy
\begin{equation}
|\delta\lambda|\geq\frac{\Delta O}{\big|\frac{\partial\langle O(\lambda)\rangle}{\partial\lambda}\big|}
\label{sens2}
\end{equation}
to be experimentally detectable~\cite{sensitivity}. The value $\delta\lambda'$ that saturates the inequality is called the observable's \textit{sensitivity}.

The sensitivity is bounded from below by the QCRB~\cite{paris},
\begin{equation}
|\delta\lambda'|\geq\frac{1}{\sqrt{MF_Q(\lambda)}}.
\label{qcrb1}
\end{equation}
Here, $M$ is the number of measurements and $F_Q(\lambda)$ is the quantum Fisher information. If a state $\rho_\lambda$ depends on the parameter $\lambda$ through some quantum operation $\rho_\lambda=\sum_kM_k(\lambda)\rho_0M_k^\dagger(\lambda)$, the quantum Fisher information can be written as~\cite{paris}
\begin{equation}
F_Q(\lambda)=2\sum_{n,m}\frac{|\langle\psi_n|\partial_\lambda\rho_\lambda|\psi_m\rangle|^2}{p_n+p_m},
\label{qcrb2}
\end{equation}
where $\{|\psi_n\rangle\}$ is the eigenbasis of the output state $\rho_\lambda$ with the corresponding eigenvalues $p_n$, and
\begin{equation}
\partial_\lambda\rho_\lambda=\sum_k\Bigg[\frac{\partial M_k(\lambda)}{\partial\lambda}\rho_0M_k^\dagger(\lambda)+M_k(\lambda)\rho_0\frac{\partial M_k^\dagger(\lambda)}{\partial\lambda}\Bigg],
\label{qcrb3}
\end{equation}
with also $\rho_0$ written in its eigenbasis.

\subsection{\label{oqs}Open quantum systems and non-Markovianity}

A quantum system $\mathcal{Q_S}$ is said to be open if it interacts with some other system, dubbed its environment $\mathcal{E}$~\cite{Breuer2007}. As a consequence, information carried by $\mathcal{Q_S}$ typically transforms into correlations between $\mathcal{Q_S}$ and $\mathcal{E}$. Trace distance $D_\text{tr}(\rho_1,\rho_2)=\frac{1}{2}\text{tr}|\rho_1-\rho_2|$ provides an intuitive way to characterize this information flow, as it is directly proportional to the distinguishability of two (open-system) states $\rho_1$ and $\rho_2$~\cite{chuang}; $\mathcal{Q_S}$ carries less information if $\rho_1$ and $\rho_2$ become less distinguishable under some environment-induced mapping $\rho\mapsto\Phi(\rho)$.

The dynamics of open quantum systems is often described by completely positive and trace-preserving (CPTP) maps, or \textit{channels} $\Phi$, under which the trace distance can never increase, i.e., $D_\text{tr}\big(\Phi(\rho_1),\Phi(\rho_2)\big)\leq D_\text{tr}(\rho_1,\rho_2)$~\cite{chuang}. However, because the total system is closed, the open system $\mathcal{Q_S}$ may sometimes regain some of the previously lost information. In such situations the trace distance can temporally increase, though never exceeding its initial value. According to the Breuer, Laine, and Piilo (BLP) definition, such nonmonotonic behavior of trace distance indicates non-Markovian dynamics and the presence of memory effects~\cite{blp}.

We note that there is no sole, universally agreed definition of quantum (non-)Markovianity~\cite{blp,nm_1,nm_3,nm_4,teittinen,budini,nm_5,nm_6,modi}. Still, the revivals of trace distance coincide with many other indicators of non-Markovianity~\cite{teittinen}. Most notably---and in the case of single-qubit dephasing, which we will be dealing with---the increase of trace distance coincides with the violation of completely positive (CP) divisibility~\cite{nm_1}.

The BLP-measure of quantum non-Markovianity is more formally defined as
\begin{equation}
\mathcal{N}(\Phi)=\max_{\rho_{1,2}(0)}\int_{\sigma>0}dt\sigma\big(t,\rho_{1,2}(0)\big),
\label{blp}
\end{equation}
where $\sigma(t,\rho_{1,2}(0)\big)=\frac{d}{dt}D_\text{tr}\big(\rho_1(t),\rho_2(t)\big)$~\cite{blp}. Hence, the BLP-measure simply equals the total increase of trace distance upon the whole time evolution induced by the channel $\Phi$, maximized with respect to the initial state pair $\rho_{1,2}(0)$. With single-qubit systems in dephasing channels, this state pair is $(|0\rangle\pm|1\rangle)/\sqrt{2}$ and the trace distance becomes the absolute value of the \textit{decoherence function} $\langle0|\rho_{1(2)}(t)|1\rangle/\langle0|\rho_{1(2)}(0)|1\rangle$~\cite{lin_opt_6}. In this work, we are interested in the sensitivity of this function and non-Markovian memory effects.

\section{\label{protocol}Protocol}

\subsection{General case}

\begin{figure*}[t!]
\includegraphics[width=.8\textwidth]{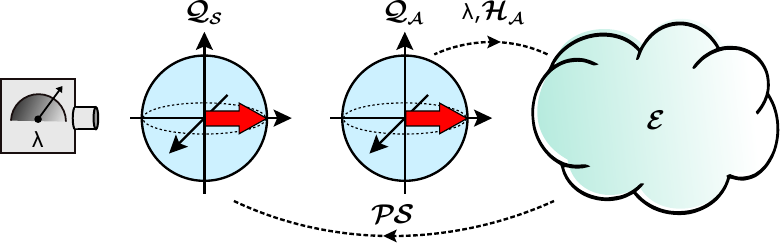}
\caption{
Schematic picture of the parameter estimation protocol. First, we let the ancillary qubit $\mathcal{Q_A}$ interact with the environment $\mathcal{E}$, encoding the parameter of interest $\lambda$ into their joint state. We also apply $\mathcal{H_A}$ to $\mathcal{Q_A}$. Then, we let the system qubit $\mathcal{Q_S}$ interact with $\mathcal{E}$ as well. We post-select ($\mathcal{PS}$) the subchannel with information backflow (the arrows point to the direction of information flow). Finally, measuring $\mathcal{Q_S}$ allows us to estimate $\lambda$ with higher precision than measuring $\mathcal{Q_A}$ directly.
}
\label{scheme}
\end{figure*}

Our parameter estimation protocol goes as follows (see Fig.~\ref{scheme} for a schematic illustration). First, we prepare an ancillary qubit $\mathcal{Q_A}$ in the balanced superposition state $|\psi_\mathcal{A}\rangle=(|0_\mathcal{A}\rangle+|1_\mathcal{A}\rangle)/\sqrt{2}$. We then let $\mathcal{Q_A}$ interact with an environment $\mathcal{E}$, encoding the parameter of interest $\lambda$ into their joint state. We assume that the interaction Hamiltonian is of the dephasing form $H_\mathcal{AE}=-\sum_{n=0,1}\lambda_n|n_\mathcal{A}\rangle\langle n_\mathcal{A}|\otimes\hat{E}$~\cite{Breuer2007}, with $\lambda\coloneqq\lambda_0-\lambda_1$. The joint state evolves as
\begin{equation}
    |\Psi_\mathcal{AE}(\tau)\rangle=\frac{1}{\sqrt{2}}\Big(|0_\mathcal{A}\rangle e^{i\hat{E}\lambda_0\tau}+|1_\mathcal{A}\rangle e^{i\hat{E}\lambda_1\tau}\Big)|\psi_\mathcal{E}\rangle,
\end{equation}
where $|\psi_\mathcal{E}\rangle$ is the initial state of the environment and we have set $\hbar=1$. To allow information backflow later, we mix the phase factors with the Hadamard operator $\mathcal{H_A}$.

The protocol proceeds with the system qubit $Q_S$, which we also couple with $\mathcal{E}$. Preparing it similarly to $Q_\mathcal{A}$ and assuming the same type of interaction---this time depending on $\nu\coloneqq\nu_0-\nu_1$---the joint state of all the three subsystems (system, ancilla, environment) reads
\begin{equation}
    \begin{split}
        |\Psi_\mathcal{SAE}(t)\rangle=
        \frac{1}{2}\Big[&|0_\mathcal{S}\rangle|0_\mathcal{A}\rangle e^{i\hat{E}\nu_0t}\big(e^{i\hat{E}\lambda_0\tau}+e^{i\hat{E}\lambda_1\tau}\big)\\
        +&|1_\mathcal{S}\rangle|0_\mathcal{A}\rangle e^{i\hat{E}\nu_1t}\big(e^{i\hat{E}\lambda_0\tau}+e^{i\hat{E}\lambda_1\tau}\big)\\
        +&|0_\mathcal{S}\rangle|1_\mathcal{A}\rangle e^{i\hat{E}\nu_0t}\big(e^{i\hat{E}\lambda_0\tau}-e^{i\hat{E}\lambda_1\tau}\big)\\
        +&|1_\mathcal{S}\rangle|1_\mathcal{A}\rangle e^{i\hat{E}\nu_1t}\big(e^{i\hat{E}\lambda_0\tau}-e^{i\hat{E}\lambda_1\tau}\big)\Big]|\psi_\mathcal{E}\rangle.
    \end{split}
\label{eq:psi_out}
\end{equation}
Note that here we treat the first interaction time $\tau$ as a fixed preparation parameter and $t$ as the actual interaction time that starts running once the state preparation is complete.

Partial-tracing over $\mathcal{Q_A}$ and $\mathcal{E}$, the state of $\mathcal{Q_S}$ becomes
\begin{equation}
    \rho_\mathcal{S}(t)=\frac{1}{2}\begin{pmatrix}
        1 & \kappa(t)\\
        \kappa(t)^* & 1
    \end{pmatrix},
\end{equation}
where
\begin{equation}
    \kappa(t)=\langle\psi_\mathcal{E}|e^{i\hat{E}\nu t}|\psi_\mathcal{E}\rangle
\end{equation}
is the decoherence function. Writing $|\psi_\mathcal{E}\rangle$ in the eigenbasis of $\hat{E}$, i.e., $|\psi_\mathcal{E}\rangle=\int d\epsilon g(\epsilon)|\epsilon\rangle$ with $\hat{E}|\epsilon\rangle=\epsilon|\epsilon\rangle$, we notice that $\kappa(t)$ is actually the characteristic function of $\epsilon$~\cite{Feller1971},
\begin{equation}
    \kappa(t)=\int d\epsilon|g(\epsilon)|^2e^{i\epsilon\nu t}.
\end{equation}
For many smooth and unimodal probability distributions $|g(\epsilon)|^2$, $|\kappa(t)|$ shows a monotonically decaying, Markovian-like behavior. Furthermore, $|\kappa(t)|$ is independent from $\lambda$. However, we can also write $\rho_\mathcal{S}(t)$ as the convex combination
\begin{equation}
    \rho_\mathcal{S}(t)=P_0\Phi_0^t\big(\rho_\mathcal{S}(0)\big)+P_1\Phi_1^t\big(\rho_\mathcal{S}(0)\big),
\end{equation}
where
\begin{equation}
    P_n=\frac{1}{2}\Big[1+(-1)^n\int d\epsilon|g(\epsilon)|^2\cos(\epsilon\lambda\tau)\Big]
    \label{P_n}
\end{equation}
is the probability of measuring $\mathcal{Q_A}$ in the state $|n_\mathcal{A}\rangle$ and
\begin{equation}
    \Phi_n^t\big(\rho_\mathcal{S}(0)\big)=\frac{1}{2}
    \begin{pmatrix}
        1 & \kappa_n(t) \\
        \kappa_n(t)^* & 1
    \end{pmatrix}
    \label{rho_n}
\end{equation}
is the corresponding quantum channel with the conditional decoherence function
\begin{equation}
\begin{split}
    \kappa_n(t)=\frac{1}{4P_n}\Big[2\kappa(t)&+(-1)^n\kappa(t+\lambda\tau/\nu)\\
    &+(-1)^n\kappa(t-\lambda\tau/\nu)\Big].
\end{split}
\end{equation}

While $\lambda$ is more commonly estimated through $P_n$, now the decoherence function depends on $\lambda$ too. Consisting of three peaks at $t=0$ and $t=\pm\lambda\tau/\nu$ as well as the interfering cross-terms, $|\kappa_n(t)|$ can exhibit richer and more sensitive behavior than $P_n$. In particular---depending on the ultrasensitive interference terms---the coherences of $\rho_n(t)$ can temporally revive around $t=|\lambda\tau/\nu|$, manifesting non-Markovian memory effects. Once we fix the physical system in the following sections, it becomes much clearer how sensitive these memory effects and the information backflow can really be.

In practice, the BLP-non-Markovianity could be evaluated by tomographing the post-selected $\mathcal{Q_S}$ at different interaction times $t$. Writing Eq.~\eqref{rho_n} in the Bloch representation, $\rho_n(t)=\frac{1}{2}\sum_{k=0}^3r_k(t)\sigma_k$ [with $r_0(t)=1$ and $\sigma_0=\mathds{1}$], it is straightforward to verify that $|\kappa_n(t)|=\sqrt{\frac{r_1(t)^2+r_2(t)^2}{1-r_3(t)^2}}$, where $r_k(t)$ are the Bloch-vector components being actually measured. $\mathcal{N}(\Phi_n^t)$ is then obtained by subtracting the local minimum of $|\kappa_n(t)|$ from its local maximum. If there are none, the protocol fails and we should measure $|\kappa_{n\oplus1}(t)|$ or $P_n$ instead.

\begin{figure}[t!]
\includegraphics[width=\linewidth]{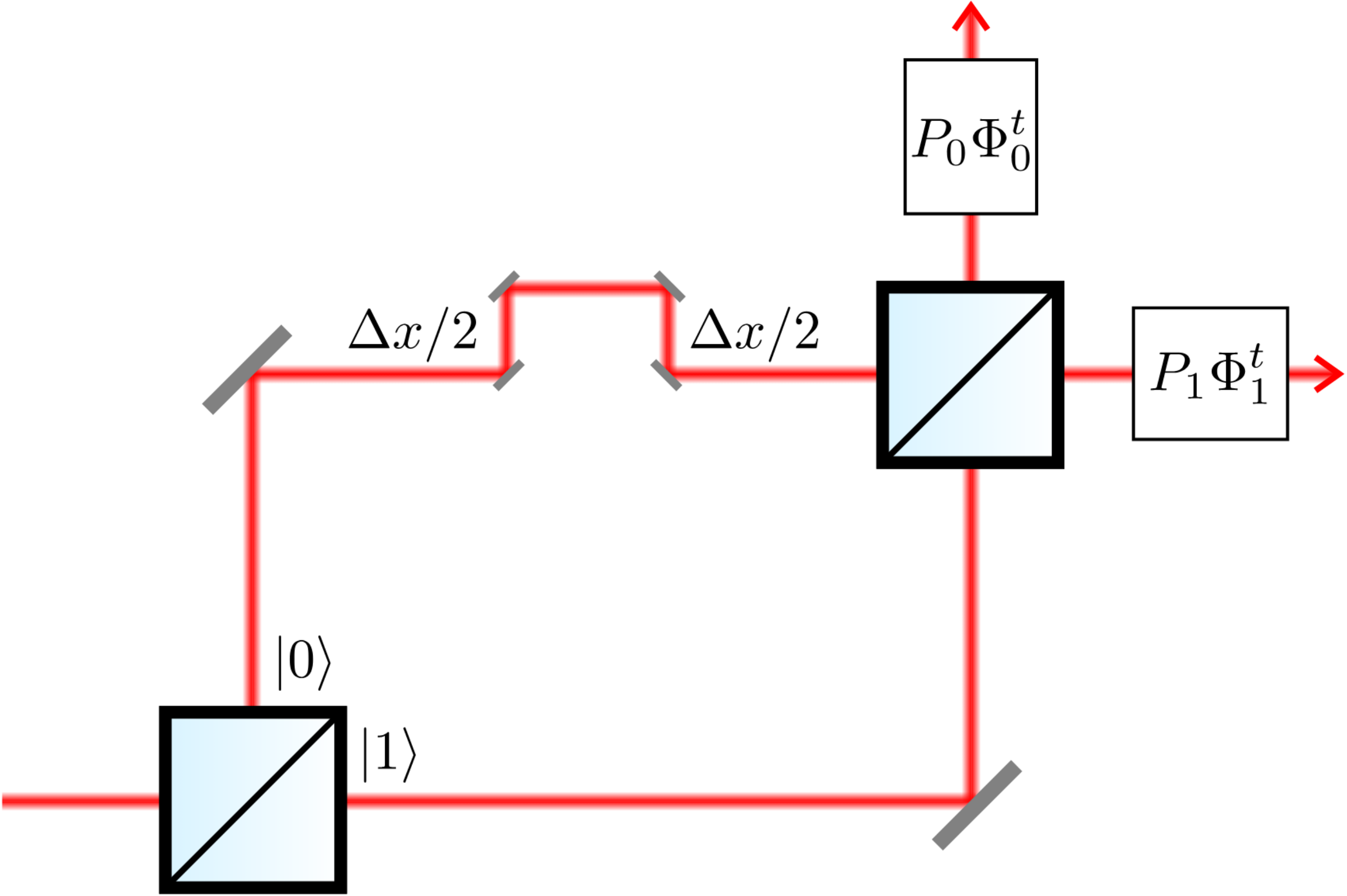}
\caption{
The linear optical setup. The ancilla $\mathcal{Q_A}$ (path) and environment $\mathcal{E}$ (frequency) are first coupled via $\Delta x\coloneqq x_0-x_1$. The channels $\Phi_0^t$ and $\Phi_1^t$ of the open system $\mathcal{Q_S}$ (polarization) are then realized with birefringent crystals, e.g., quartz.
}
\label{setup}
\end{figure}

\subsection{Linear optical implementation}

The protocol can be realized in linear optical framework (see Fig.~\ref{setup}). Here, the polarization degree of freedom of single photons is interpreted as the open system $\mathcal{Q_S}$, their path as the ancillary qubit $\mathcal{Q_A}$, and their frequency as the environment $\mathcal{E}$. The parameter we are interested in is the path difference of an unbalanced Mach-Zehnder interferometer, which couples the path and frequency. The first beam splitter transforms the input state $|0_\mathcal{A}\rangle$  to $(|0_\mathcal{A}\rangle+|1_\mathcal{A})/\sqrt{2}$, while the second one mixes the channels. The polarization channels can then be realized by guiding the photons exiting the interferometer into birefringent crystals. The interaction time $t$ can be controlled by varying the thickness $d=ct$ of these crystals, with $c$ being the speed of light in vacuum.

The path-frequency interaction Hamiltonian is
\begin{equation}
    H_\mathcal{AE}=-\frac{1}{c\tau}(x_0|0\rangle\langle0|+x_1|1\rangle\langle1|)\otimes\int df2\pi f|f\rangle\langle f|,
\end{equation}
where $x_n$ is the length of path $n=0,1$. The polarization-frequency Hamiltonian, in turn, reads
\begin{equation}
    H_\mathcal{SE}=-(n_H|H\rangle\langle H|+n_V|V\rangle\langle V|)\otimes\int df2\pi f|f\rangle\langle f|.
\end{equation}
Here, $n_{H(V)}$ is the refractive index of horizontal (vertical) polarization. Note that we have dropped the subscripts $\mathcal{A}$ and $\mathcal{S}$ from the qubit states due to the already distinct notation.

Using a Gaussian spectrum with the central frequency $\mu$ and standard deviation $\sigma$,
\begin{equation}
|g(f)|^2=\frac{1}{\sqrt{2\pi\sigma^2}}e^{-\frac{1}{2}\big(\frac{f-\mu}{\sigma}\big)^2},
\end{equation}
the decoherence function of polarization, prior to post-selection, becomes
\begin{equation}
\kappa(t)=e^{i2\pi\mu\Delta nt-\frac{1}{2}(2\pi\sigma\Delta nt)^2}.
\label{gaussian_deco}
\end{equation}
$\Delta n=n_H-n_V$ is the birefringence of the crystals. As discussed earlier, $|\kappa(t)|$ decays monotonically and does not depend on $\Delta x$. However, if we post-select just one of the paths $n$ instead of averaging over them, we get the pathwise, conditional decoherence function
\begin{equation}
    \begin{split}
        \kappa_n(t)=\frac{1}{4P_n}\Big[2\kappa(t)+&(-1)^n\kappa\big(t+\Delta x/(c\Delta n)\big)\\+&(-1)^n\kappa\big(t-\Delta x/(c\Delta n)\big)\Big],
    \end{split}
\end{equation}
where the probability to detect a photon on path $n$ reads
\begin{equation}
    P_n=\frac{1}{2}\Big\{1+(-1)^n\exp\Big[-\frac{1}{2}(2\pi\sigma\Delta x/c)^2\Big]\cos(2\pi\Delta x/\lambda_0)\Big\}.
\end{equation}
$\lambda_0=c/\mu$ is the central wavelength of the photon source.

Physically, $\kappa_n(t)$ describes how the orthogonal polarizations combine in the second beam splitter and interfere afterwards in the birefringent crystals. That is, $\kappa(t)$ describes how $H$ and $V$ coming from the same path behave, whereas $\kappa\big(t\pm\Delta x/(c\Delta n)\big)$ describes the dynamics of $H$ and $V$ coming from different paths inside the interferometer.

%For example, $|\kappa(t)|=\text{exp}[-(2\pi\sigma\Delta nt)^2/2]$ decreases monotonically, making the total channel $\Phi^t$ in Eq.~\eqref{tot_channel} Markovian. However, the channels $\Phi_0^t$ and $\Phi_1^t$ can both be Markovian (M+M), non-Markovian (nM+nM), or one Markovian, the other non-Markovian (M+nM or nM+M), as we will soon see. Here, the path difference $\Delta x$---and so, the concurrence of the environment state---acts as the source of non-Markovianity (see Appendix A for details).

%Since we are essentially interested in $|\kappa_j(t)|$ only, the BLP-non-Markovianity can be measured by simple polarization tomography and using just one initial state, say, $|+\rangle$. Writing Eq.~\eqref{rho_j} in terms of the Stokes parameters and Pauli matrices, $\rho_j(t)=\frac{1}{2}\sum_{n=0}^3S_n(t)\sigma_n$ [with $S_0(t)=1$ and $\sigma_0=\mathds{1}$], it is straightforward to verify that $|\kappa_j(t)|=\sqrt{\frac{S_1(t)^2+S_2(t)^2}{1-S_3(t)^2}}$. The interaction time $t$ can be controlled, e.g., by stacking quartz plates one after another, with their total thickness $x$ satisfying $x=ct$. At each quartz plate combination, we measure the Stokes parameters and evaluate $|\kappa_j(t)|$. Finally, after having repeated polarization tomography for several quartz plate combinations, we obtain $\mathcal{N}(\Phi_j^t)$ by subtracting the local minimum of $|\kappa_j(t)|$ from its local maximum. If there are none, $\mathcal{N}(\Phi_j^t)=0$ and the protocol fails.

\begin{figure}[b!]
\includegraphics[width=\linewidth]{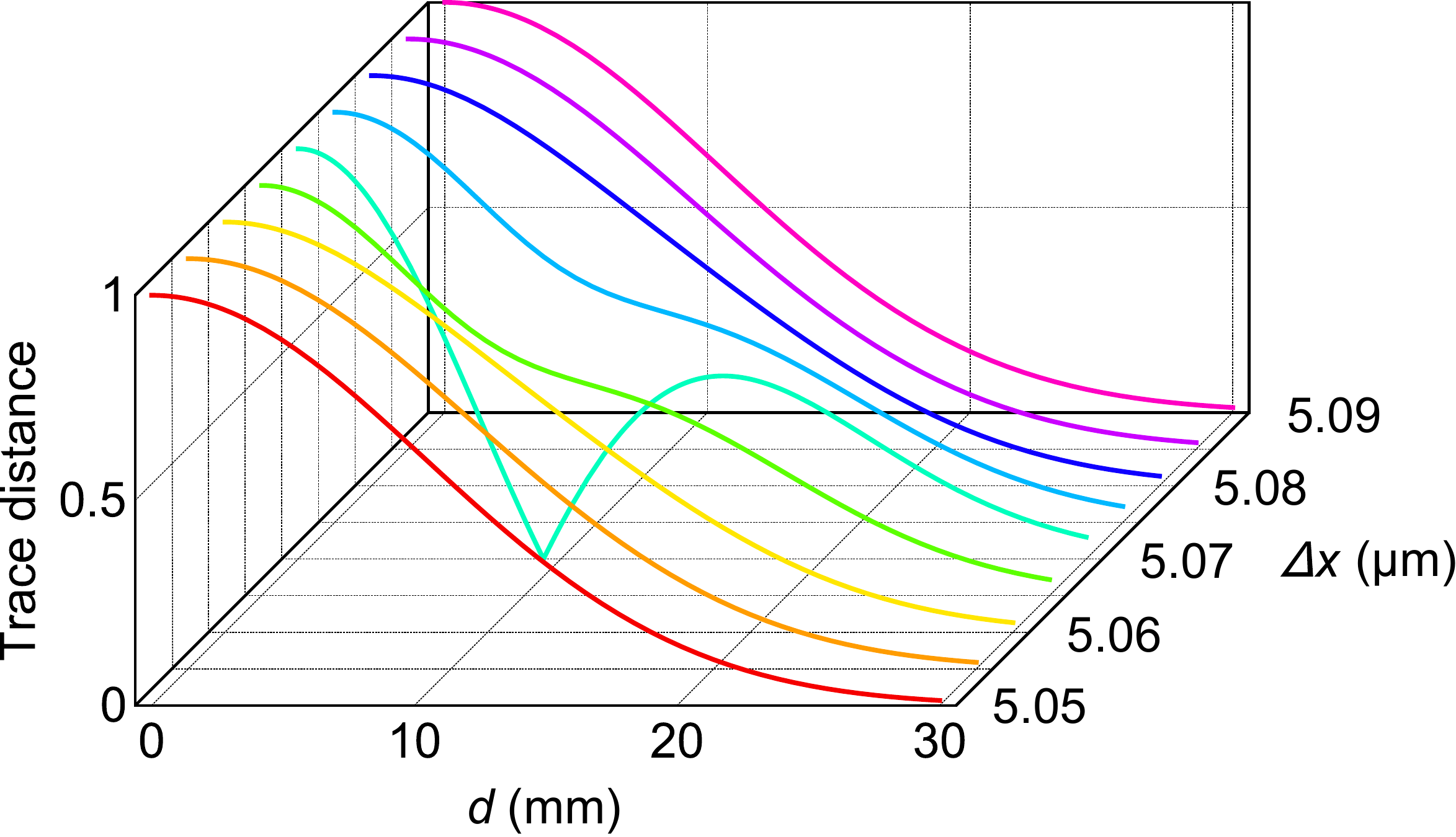}
\caption{
Trace-distance dynamics of the initial state pair $|\pm\rangle$ as a function of the thickness of quartz, shown for different values of $\Delta x$. $\lambda_0=780$ nm, $\sigma=5.68\times10^{11}\text{ Hz}$, and $\Delta n=0.009$.
}
\label{dtr}
\end{figure}

\begin{figure*}[t!]
\includegraphics[width=\textwidth]{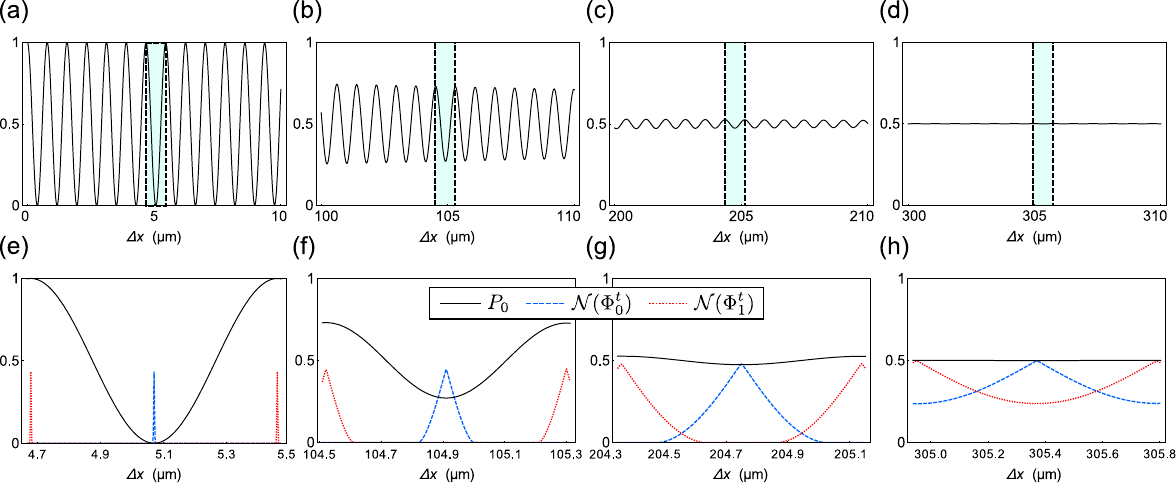}
\caption{
Path probabilities $P_0$ (black, solid) and BLP-non-Markovianities $\mathcal{N}(\Phi_0^t)$ (blue, dashed) and $\mathcal{N}(\Phi_1^t)$ (red, dotted) as functions of the path difference $\Delta x$. The panels (e)--(h) are magnifications of the highlighted regions in (a)--(d), respectively. $\lambda_0=780$ nm, $\sigma=5.68\times10^{11}\text{ Hz}$, and $\Delta n=0.009$.
}
\label{p_blp}
\end{figure*}

\section{\label{results}Simulation results}

\subsection{General remarks}

To visualize the memory effects and their dependency on $\Delta x$, we have plotted the trace-distance dynamics of the state pair $\Phi_0^t(|\pm\rangle\langle\pm|)$, with $|\pm\rangle=(|H\rangle\pm|V\rangle)/\sqrt{2}$, in Fig.~\ref{dtr} for different values of $\Delta x$. With the example parameters, we see a sudden emergence of non-Markovian memory effects just $\pm5$ nm around $\Delta x=5.07$ $\upmu$m, which indicates high sensitivity. However, in Fig.~\ref{dtr} we also have $P_0<0.01$, which means that the memory effects can be demanding to measure due to low signal.

Fig.~\ref{p_blp}---with Fig.~\ref{p_blp}(e) corresponding to Fig.~\ref{dtr}---better illustrates the interplay of $P_n$ and $\mathcal{N}(\Phi_n^t)$. The path probabilities $P_n$ oscillate with a frequency determined by $\mu$ and approach 1/2 with a damping rate determined by $\sigma$. Interestingly, the BLP-non-Markovianities $\mathcal{N}(\Phi_n^t)$---which are evaluated numerically in Fig.~\ref{p_blp}---peak at the local minima of $P_n$. With small values of $\Delta x$, the peaks are very narrow and do not overlap. That is, both channels can be Markovian simultaneously or only one of them non-Markovian. As the peaks become wider and wider with increasing $\Delta x$, they both approach the constant value of 1/2 and, when overlapping, make both of the channels non-Markovian. Although the $\mathcal{N}(\Phi_n^t)$-peaks become wider, they are always much narrower than the corresponding $P_n$-dips. Thus, the memory effects would seem to provide a much more sensitive method to estimate $\Delta x$.

\subsection{Sensitivity of BLP-non-Markovianity}

To evaluate the sensitivity of BLP-non-Markovianity in a more formal fashion, we first need to construct the corresponding observable. Approximating the maximum value of reviving trace distance by 0.5 (error $<$ 0.055), the degree of non-Markovianity is fully governed by the trace distance's local minimum (cf. Fig.~\ref{dtr}). Hence, in this special case, we can approximate the BLP-measure with the help of a single, properly chosen positive operator-valued measure (POVM). Defining $\Pi_{n,\text{min}}:=|\phi_{n,\text{min}}\rangle\langle\phi_{n,\text{min}}|$, where
\begin{equation}
|\phi_{n,\text{min}}\rangle=\frac{1}{\sqrt{2}}\Big[|H\rangle+e^{-i\arg\big(\kappa_n(t_\text{min})\big)}|V\rangle\Big]
\label{povm}
\end{equation}
and $t_\text{min}$ is the point of time of the trace distance's local minimum, $t_\text{min}=\inf\Big\{t:\frac{d|\kappa_n(t)|}{dt}>0\Big\}$, we can write
\begin{align}
\mathcal{N}(\Phi^t_n)&=|\kappa_n(t_\text{max})|-|\kappa_n(t_\text{min})|\\
&\approx\frac{1}{2}-\big(2\text{tr}[\Pi_{n,\text{min}}\rho_n(t_\text{min})]-1\big)\\
&=\frac{3}{2}-2\langle\phi_{n,\text{min}}|\rho_n(t_\text{min})|\phi_{n,\text{min}}\rangle.
\label{appr_BLP}
\end{align}
Here, we used $\rho_n(0)=|+\rangle\langle+|$. The greater the path difference, the smaller the error of this approximation.

The above approximation means that full process tomography might not be needed. Furthermore, since $t_\text{min}$ depends on $\Delta x$, the measurement scheme requires some \textit{a priori} knowledge about the parameter of interest, e.g., a rough estimate. Nonetheless, it is interesting to compare the sensitivity of $\mathcal{N}(\Phi_n^t)$ against that of $P_n$ and the respective QCRBs. %Namely, the QCRB can usually be saturated with parameter-dependent measurements~\cite{barndorff,locc}, and this is what we find as well.
For $P_n$, the POVM element needed to calculate the sensitivity is $\mathds{1}\otimes\mathds{1}\otimes|n\rangle\langle n|$, while the state of the system is $|\Psi_\mathcal{SAE}(t)\rangle$ [see Eq.~\eqref{eq:psi_out}]. The QCRB of $\mathcal{N}(\Phi_n^t)$ can be calculated using Eqs.~\eqref{qcrb1}--\eqref{qcrb3} and the Kraus operators
\begin{equation}
M_\pm(t_\text{min})=\sqrt{\frac{1\pm|\kappa_n(t_\text{min})|}{2}}
\begin{pmatrix}
\pm\kappa_n(t_\text{min})/|\kappa_n(t_\text{min})| & 0\\
0 & 1\end{pmatrix}
\label{Kraus}
\end{equation}
acting on the initial state $|+\rangle\langle+|$. For $P_n$, we simply replace $\kappa_n(t_\text{min})$ by $\kappa\big(\Delta x/(c\Delta n)\big)$.

\begin{figure}[t!]
\includegraphics[width=\linewidth]{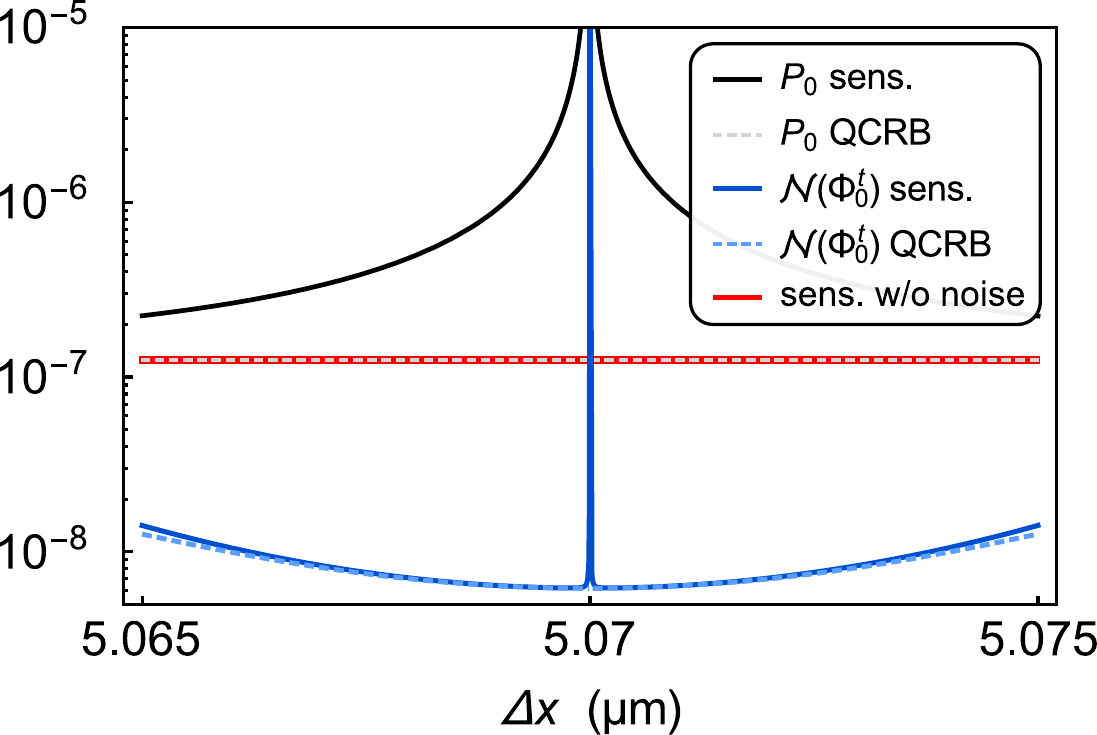}
\caption{
Theoretical sensitivities and QCRBs. Black, solid: sensitivity of $P_0$ to $\Delta x$. Gray, dashed: the corresponding QCRB for $M=1$. Blue, solid: sensitivity of $\mathcal{N}(\Phi_0^t)$ to $\Delta x$. Light blue, dashed: the corresponding QCRB for $M=1$. Red, solid: sensitivity of $P_0$ to $\Delta x$ without noise ($\sigma=0$). $\lambda_0=780$ nm, $\sigma=5.68\times10^{11}\text{ Hz}$, and $\Delta n=0.009$.
}
\label{saturation}
\end{figure}

Fig.~\ref{saturation} shows the sensitivities and QCRBs of $P_0$ and $\mathcal{N}(\Phi_0^t)$ as functions of $\Delta x$, with the same example parameters as before. Fig.~\ref{saturation} thus corresponds to Figs.~\ref{dtr} and~\ref{p_blp}(e). We recall that $\Delta x$ is the unperturbed path-difference of a Mach-Zehnder interferometer, while the sensitivities indicate how large perturbations of $\Delta x$ can be unambiguously detected by measuring $P_0$ or $\mathcal{N}(\Phi_0^t)$. Changes below the sensitivity curves could also be attributed to the quantum uncertainty of the measured observable.

As expected, the sensitivity of $\mathcal{N}(\Phi_0^t)$ clearly beats that of $P_0$. In fact, the closer $\Delta x$ is to the singular value of 5.07~$\upmu$m, the larger the difference, even three orders of magnitude. Furthermore, the sensitivity of $\mathcal{N}(\Phi_0^t)$ seems to reach its QCRB near $\Delta x\approx5.07\text{ }\upmu$m. The QCRB of $P_0$, in turn, is beaten by approximately one order of magnitude. For smaller values of $\Delta x$ and $\lambda_0$, one might expect even smaller sensitivities.

Finally, we have also plotted the noiseless sensitivity of $P_0$ in Fig.~\ref{saturation}. Here, $\sigma=0$ and there is no dephasing; the path state before the second beam splitter is the pure state $\big(|0\rangle+e^{-i2\pi\mu\Delta x/c}|1\rangle\big)/\sqrt{2}$. Indeed, $\mu$ plays the same role as the number of probes in the multiparticle parameter estimation scheme~\cite{Huelga1997}. Interestingly, the sensitivity of the dephasing-free case coincides with the QCRB of $P_0$ \textit{with} dephasing, so our non-Markovian protocol beats the former as well.

While the enhanced sensitivities might be attributed to mere post-selection---usually, the quantum Fisher information is inversely proportional to the path probability~\cite{post}---the narrow BLP-peaks in Fig.~\ref{p_blp} arise from the combination of post-selection \textit{and} dephasing. Hence, our protocol demonstrates how dephasing and non-Markovianity can serve as useful resources when properly engineered, in some cases even outperforming the noiseless scenario.

\iffalse
In Fig.~\ref{saturation}, we have plotted the sensitivity of $\Pi_{0,\text{min}}$ and, for reference, $P_0$ with respect to $\Delta x$. With $P_0$, the POVM element is $\mathds{1}\otimes\mathds{1}\otimes|0\rangle\langle0|$ and the state of the system $|\Psi_\text{out}\rangle$ [Eq.~\eqref{evolved_total}]. Fig.~\ref{saturation} also shows the QCRB with $M=1$. We calculated the QCRB using Eqs.~\eqref{qcrb1}--\eqref{qcrb3} and the Kraus operators
\begin{equation}
M_\pm(t_\text{min})=\sqrt{\frac{1\pm|\kappa_0(t_\text{min})|}{2}}
\begin{pmatrix}
\pm\kappa_0(t_\text{min})/|\kappa_0(t_\text{min})| & 0\\
0 & 1\end{pmatrix}
\label{Kraus}
\end{equation}
acting on the initial polarization state $\rho_0(0)=|+\rangle\langle+|$.
\fi

\section{\label{discussion}Conclusions and discussion}

In this work, we have introduced a non-Markovianity-based parameter estimation protocol and analyzed its performance, using a linear optical implementation as a detailed example. In the protocol, two qubits interact sequentially with a common environment that induces controllable dephasing. The resulting open-system evolution is described by a convex combination of quantum channels that can exhibit either Markovian or non-Markovian behavior. By selectively analyzing one of these channels, the parameter of interest---path difference in the photonic setup---can be estimated with high precision. When approximated with a single POVM, the BLP-based sensitivity not only reaches its own QCRB, it breaks the QCRB of the Markovian case as well as the noiseless sensitivity, proving decoherence and non-Markovianity as useful resources. At the same time, our protocol reveals new aspects of quantum interference.

We reach such high precisions, because we go beyond the standard parameter estimation scheme where one extracts information on parameter $\lambda$ by measurements on a fixed state $\rho_\lambda$. Here, we instead focus on the global properties of the channel $\Phi_n^t$. Although it is hardly surprising that overall dynamics can contain more information than a quantum state at single point of time, this raises an important question for future research: What is the precision limit for quantum processes? Previous works related to noisy parameter estimation have treated channels as snapshots of the open-system dynamics, giving interaction time dependent bounds~\cite{Escher2011,dyn_qcrb_2,dyn_qcrb_3}, but here we mean the overarching dynamics. One potential starting point for such a theory could be \textit{process positive operator-valued measures} (PPOVMs)~\cite{ppovm}.

Although our results have exciting implications on parameter estimation, the protocol also has its shortcomings. With small path differences the memory effects---though extremely sensitive---become less probable to appear, and if they do, the photon counts are equally low. Also, full process tomography is much more tedious than single projective measurements. We thus anticipate that an optimal parameter estimation protocol might combine more traditional methods with ours. For example, one could first use more conventional methods to determine if we are in the non-Markovian regime and then employ our protocol to gain better precision.

The proposed protocol is not limited to photonic implementations. It can be realized in a variety of physical platforms where controllable dephasing and memory effects are accessible. In trapped-ion systems, electronic qubits may couple to vibrational modes that act as a structured reservoir, while high-fidelity single-qubit gates and heralded read-out make post-selection and coherence measurements feasible~\cite{Ge2019}. Similarly, superconducting qubits, semiconductor quantum dots, nitrogen-vacancy centers in diamond, and cold atoms in optical lattices could all host analogous non-Markovianity-based estimation protocols.

%\vspace{10pt}

\section*{Acknowledgments}

The Author acknowledges the financial support from the Magnus Ehrnrooth Foundation and the fruitful discussions with J. Piilo, T. Kuusela, and L. Santos.

\end{document}